\documentclass[conf]{IEEEtran}
\usepackage{amsmath,amssymb}
\usepackage{graphicx}
\usepackage{mathrsfs}
\usepackage{tikz}
\usepackage{booktabs}
\usepackage{floatflt}
\usepackage{enumerate}
\usepackage{psfrag}	
\usepackage{array}
\usepackage{multirow,hhline}
\usepackage{exscale}
\usepackage{color}
\usepackage{colortbl}
\usepackage{epsfig,subfigure}
\usepackage{cite}
\usepackage{amsthm}
\usepackage{dsfont}
\usepackage{nicefrac}

\newcommand{\alert}[1]{\textcolor{black}{#1}}
\renewcommand{\vec}[1]{\ensuremath{\mathbf{#1}}}
\newtheorem{theorem}{Theorem}

\newtheorem{remark}{Remark}

\newcommand{\y}[0]{\ensuremath{\mathbf{y}}}
\newcommand{\z}[0]{\ensuremath{\mathbf{z}}}
\renewcommand{\H}[0]{\ensuremath{\mathbf{H}}}
\newcommand{\x}[0]{\ensuremath{\mathbf{x}}}
\newcommand{\V}[0]{\ensuremath{\mathbf{V}}}

\newcommand{\N}[0]{\ensuremath{\mathbf{N}}}
\newcommand{\T}[0]{\ensuremath{\mathbf{T}}}
\newcommand{\D}[0]{\ensuremath{\mathbf{D}}}
\renewcommand{\u}[0]{\ensuremath{\mathbf{u}}}
\newcommand{\w}[0]{\ensuremath{\mathbf{w}}}

\DeclareMathOperator{\spn}{span}

\begin{document}


\IEEEoverridecommandlockouts
\title{The Degrees of Freedom of the MIMO Y-channel}
\author{
\IEEEauthorblockN{Anas Chaaban, Karlheinz Ochs, and Aydin Sezgin}\\
\IEEEauthorblockA{Chair of Communication Systems\\
Ruhr-Universit\"at Bochum (RUB), Germany\\
Email: {anas.chaaban,karlheinz.ochs,aydin.sezgin@rub.de}}\\
\thanks{%
This work is supported by the German Research Foundation, Deutsche
Forschungsgemeinschaft (DFG), Germany, under grant SE 1697/7.
}
}

\maketitle

\thispagestyle{empty}

\begin{abstract}
The degrees of freedom (DoF) of the MIMO Y-channel, a multi-way communication network consisting of 3 users and a relay, are characterized for arbitrary number of antennas. The converse is provided by cut-set bounds and novel genie-aided bounds. The achievability is shown by a scheme that uses beamforming to establish network coding on-the-fly at the relay in the uplink, and zero-forcing pre-coding in the downlink. It is shown that the network has $\min\{2M_2+2M_3,M_1+M_2+M_3,2N\}$ DoF, where $M_j$ and $N$ represent the number of antennas at user $j$ and the relay, respectively. Thus, in the extreme case where $M_1+M_2+M_3$ dominates the DoF expression and is smaller than $N$, the network has the same DoF as the MAC between the 3 users and the relay. In this case, a decode and forward strategy is optimal. In the other extreme where $2N$ dominates, the DoF of the network is twice that of the aforementioned MAC, and hence network coding is necessary. As a 
byproduct of this work, it is shown that channel output feedback from the relay to the users has no impact on the DoF of this channel.
\end{abstract}



\section{Introduction}
Multi-way relay channels have attracted extensive research attention recently, as they constitute an integral part of future networks due to the possibility of employing sophisticated communication techniques (such as network coding) in such networks to increase the spectral efficiency. 

This line of research has started with the two-way relay channel which represents the most basic multi-way relay channel. This channel was introduced in \cite{RankovWittneben}, and further analyzed in \cite{KimDevroyeMitranTarokh,GunduzTuncelNayak,NamChungLee_IT,AvestimehrSezginTse,OechteringBjelakovicSchnurrBoche}. The capacity region of this channel is known within a constant gap. Recently, the two-way two-relay channel has also been studied and its symmetric capacity approximated in \cite{SongDevroyeShaoNgo}. The multi-way relay channel with more than 2 users was also studied in \cite{GunduzYenerGoldsmithPoor,OngKellettJohnson} where each user has 1 message to be multicast to other users, and multi-pair two-way relay channels have been studied in \cite{SezginAvestimehrKhajehnejadHassibi,SezginBocheAvestimehr}.

In this paper, we consider a 3-user MIMO multi-way relay channel, where each user has two messages to be delivered to the two other users. This setup, known as the Y-channel, was introduced by Lee {\it et al.} in \cite{LeeLimChun}, where it was shown that if the relay has $N\geq\lceil{3M/2}\rceil$ antennas where $M$ is the number of antennas at each user, then the cut-set bound given by $3M$ DoF is achievable. Contrary to this case, it was shown in \cite{ChaabanSezginAvestimehr_YC_SC} that the cut-set bound is not achievable in the SISO Y-channel, and that new bounds are necessary to obtain an approximate characterization of the sum-capacity. In \cite{Jafar}, it was noted that the DoF of the general Y-channel with arbitrary number of antennas is still open. This problem is settled in this paper.

The novel contribution of the paper is a complete characterization of the DoF of the MIMO Y-channel. We show that the general MIMO Y-channel with $M_1\geq M_2\geq M_3$ antennas at user 1, 2, and 3, respectively, and with $N$ antennas at the relay, has {$\min\{2M_2+2M_3,M_1+M_2+M_3,2N\}$} DoF. To show the converse of this result, it is necessary to use genie-aided bounds similar to those in \cite{ChaabanSezginAvestimehr_YC_SC}. These novel bounds are essential for characterizing the DoF of the MIMO Y-channel, and cover the opposite case to the one considered by Lee {\it et al.} \cite{LeeLimChun}. The genie-aided bounds provide a complete characterization of the DoF of the network when combined with the cut-set bounds \cite{CoverThomas, LeeLimChun}. The achievability result is provided by signal-space alignment for network coding in the uplink \cite{LeeLimChun} and zero-forcing pre-coding in the downlink. As a byproduct, it is noted that the DoF can be achieved without using the received signal at each user 
in the encoding process. This implies that channel output feedback from the relay to the users has no impact on the DoF of this channel. 

Throughout the paper, we use bold-face lower and upper case letters for vectors and matrices, respectively, and we use $(.)^\dagger$ to denote the transpose of a matrix. $\vec{I}$ is the identity matrix, and $\vec{0}$ is the zero vector. $\x^n$ is used to denote the sequence $(\x_1,\cdots,\x_n)$. The rest of the paper is organized as follows. In Section \ref{Sec:Model}, the system model is introduced and the DoF theorem is stated. In Sections \ref{Sec:UB} and \ref{Sec:LB}, the converse and the achievability proof of the Theorem are given, respectively. Finally, we discuss the result in Section \ref{Sec:Disc}.

\section{System Model}
\label{Sec:Model}
\begin{figure}
\centering
\includegraphics[width=.8\columnwidth]{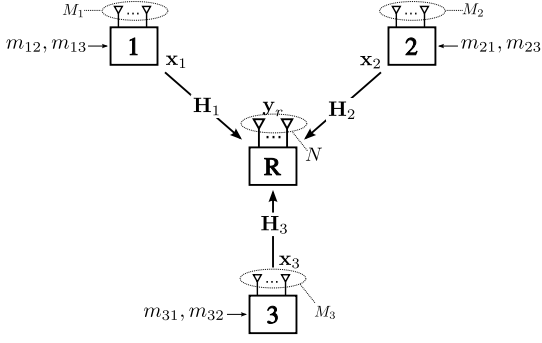}
\caption{{The MIMO Y-channel in the uplink.}}
\label{Fig:ModelU}
\end{figure}

The Y-channel consists of 3 users and a relay as shown in Fig. \ref{Fig:ModelU}, where each user wants to commmunicate with the other two users. \alert{We assume that all nodes are full-duplex and that the channels and signals are real valued. Moreover, the channels are assumed to hold the same value for the duration of the transmission.} User $j$ has messages $m_{jk}$ and $m_{jl}$ to be sent to users $k$ and $l$, with rates $R_{jk}$ and $R_{jl}$, respectively, for distinct $j,k,l\in\{1,2,3\}$. The transmit signal of user $j$ at time instant\footnote{{The time index $i$ will be dropped in the sequel unless necessary}} $i$ is an $M_j\times1$ vector {$\x_{j}(i)$}. This transmit signal is in general constructed from all the information available at user $j$ (messages and observed signals). The received signal at the relay is given by
\begin{align}
\y_{r}(i)=\H_1\x_{1}(i)+\H_2\x_{2}(i)+\H_3\x_{3}(i)+\z_{r}(i),
\end{align}
which is an $N\times1$ vector, where $\z_{r}(i)$ is an i.i.d. Gaussian noise vector $\z_r\sim\mathcal{N}(\vec{0},\vec{I})$ and $\H_j$ is the $N\times M_j$ random channel matrix from user $j$ to the relay. We assume without loss of generality that 
\begin{align}
\label{Ordering}
M_1\geq M_2\geq M_3.
\end{align}
The relay uses its observations up to time instant $i-1$ to construct $\x_{r}(i)$. The received signal at user $j$ is given by
\begin{align}
\label{ReceivedSignal}
\y_{j}(i)=\D_j\x_{r}(i)+\z_{j}(i),
\end{align}
which is an $M_j\times1$ vector, where $\z_{j}(i)$ is an i.i.d. Gaussian noise vector $\z_j\sim\mathcal{N}(\vec{0},\vec{I})$, {and $\D_j$ is the $M_j\times N$ downlink channel matrix\footnote{{For reciprocal channels, $\D_j=\H_j^\dagger$.}} from the relay to user $j$}. All nodes have a power constraint $P$.
The DoF of message $m_{jk}$ is defined as 
\begin{align}
\label{DoFDef}
d_{jk}=\lim_{\substack{P\to\infty}}\frac{R_{jk}}{\frac{1}{2}\log(P)},
\end{align}
and the sum-DoF denoted by $d_\Sigma$ is the sum of all $d_{jk}$. The following theorem states the DoF of the MIMO Y-channel.
\begin{theorem}
\label{Thm:DoF}
The DoF of the MIMO Y-channel with $M_1\geq M_2\geq M_3$ is given by
\begin{align}
\label{Eq:DoF}
{d_\Sigma=\min\{2M_2+2M_3, M_1+M_2+M_3, 2N\}.}
\end{align}
\end{theorem}

\begin{remark}
The special case of $M_1=M_2=M_3=M$ with $N\geq\left\lceil\frac{3M}{2}\right\rceil$ which was studied by Lee {\it et al.} in \cite{LeeLimChun} is covered by this theorem, where $3M$ DoF can be achieved.
\end{remark}
The first and last arguments of the $\min$ operation in \eqref{Eq:DoF} are obtained from novel genie-aided bounds which are essential for complete DoF characterization, whereas the remaining term is obtained from cut-set bounds. In the derivation of the upper bounds, it is assumed that the users use non-restricted encoders. That is, the transmit signal of each user at time $i$ depends not only on the messages, but also on the received signals at this user till time instant $i-1$. This received signal can be considered as channel output feedback, i.e., a feedback of $\y_r(1),\cdots,\y_r(i-1)$. On the other hand, the transmission scheme used to show the achievability of this theorem does not make use of the received signals at each user for encoding. This means that channel output feedback does not have any influence on the DoF of this network. The following sections are devoted for the proof of this Theorem \ref{Thm:DoF}.

\section{Upper Bounds}
\label{Sec:UB}
In this section, we prove the converse of Theorem \ref{Thm:DoF}. For this purpose, we need two types of bounds: the cut-set bounds \cite{CoverThomas}, and genie-aided bounds. We start with the cut-set bounds.
\begin{figure}
\centering
\includegraphics[width=.8\columnwidth]{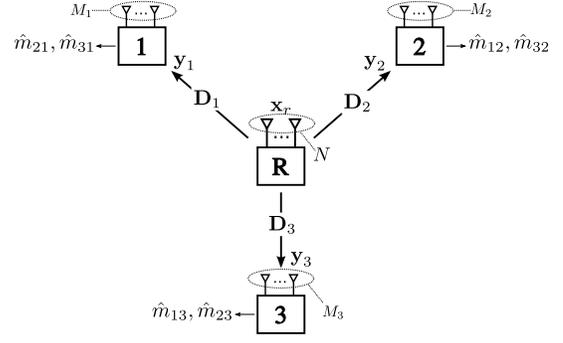}
\caption{{The MIMO Y-channel in the downlink.}}
\label{Fig:ModelD}
\end{figure}

\vspace{-.5cm}
\subsection{Cut-set Bounds}
In \cite{LeeLimChun}, Lee {\it et al.} derived cut-set bounds for the MIMO Y-channel with equal number of antennas at all users. Next, we generalize these bounds for arbitrary number of antennas. By considering information flow from user 1 to all other users and using the cut-set bounds, we can bound the rates of the corresponding messages as follows
\begin{align}
\label{CutSet1}
R_{12}+R_{13}&\leq I(\x_1;\y_2,\y_3,\y_r|\x_2,\x_3),\\
\label{CutSet2}
R_{12}+R_{13}&\leq I(\x_1,\x_r;\y_2,\y_3|\x_2,\x_3).
\end{align}
These bounds lead to the following DoF bounds
\begin{align}
\label{CutSetDof1}
d_{12}+d_{13}&\leq \min\{M_1,N\},\\
\label{CutSetDof2}
d_{12}+d_{13}&\leq \min\{N,M_2+M_3\},
\end{align}
respectively. Thus, by combining \eqref{CutSetDof1} and \eqref{CutSetDof2}, we get
\begin{align}
d_{12}+d_{13}&\leq \min\{N,M_1,M_2+M_3\}.
\end{align}
By applying the same bounding technique to users 2 and 3, we can show that the following upper bound can be obtained
\begin{align}
d_{jk}+d_{jl}&\leq \min\{N,M_j,M_k+M_l\},
\end{align}
for distinct $j,k,l\in\{1,2,3\}$. By adding the three obtained bounds (for all distinct $j,k,l$), and using {$M_1\geq M_2\geq M_3$ (see \eqref{Ordering})}, we get
\begin{align}
\label{CutSetDoFBound}
&d_\Sigma\\
&\leq \min\{N,M_1,M_2+M_3\}+\min\{N,M_2\}+\min\{N,M_3\}\nonumber.
\end{align}

\vspace{-.5cm}

\subsection{Genie-aided Bounds}
In \cite{ChaabanSezginAvestimehr_YC_SC}, genie aided bounds have been derived for the SISO Y-channel. These bounds can be extended to obtain bounds for the general MIMO case. Instead of simply extending the bounds in \cite[Lemmas 1 and 2]{ChaabanSezginAvestimehr_YC_SC}, we provide a different approach to obtain them. This approach gives an insight on how to choose the genie signals needed to derive such bounds. 

Assume that every node can obtain its messages with an arbitrarily small probability of error. Using the available information at user 2 for instance, i.e., $\y_2^n$, $m_{21}$, and $m_{23}$, user 2 can obtain $m_{12}$ and $m_{32}$ reliably. Now, what additional information should be given to user 2 to allow it to decode more messages? Note that after decoding its desired messages, user 2 knows $\y_2^n$, $m_{21}$, $m_{23}$, $m_{12}$, and $m_{32}$. In order to enable user 2 to decode more messages, we should provide it with enough side information which makes it stronger than some other user. Suppose we want to make user 2 stronger than user 3 (which has $\y_3^n$, $m_{31}$ and $m_{32}$). User 2 already has $m_{32}$, thus, if we provide it with $\y_3^n$ and $m_{31}$, we make it stronger than user 3, and hence, able to decode $m_{13}$. As a result, we can write
\begin{align}
&n(R_{12}+R_{32}+R_{13}-\varepsilon_n)\nonumber\\
&\quad \leq I(m_{12},m_{32},m_{13};\y_2^n,\y_3^n,m_{21},m_{23},m_{31})\\
&\quad \leq \sum_{i=1}^nh\left(\begin{bmatrix}\D_2\\\D_3\end{bmatrix}\x_{r}(i)+\begin{bmatrix}\z_{2}(i)\\\z_{3}(i)\end{bmatrix}\right)+n\mathcal{O}(1),
\end{align}
where $\varepsilon_n\to0$ as $n\to\infty$, and where the second step follows by using standard information-theoretic operations. {Here, $\mathcal{O}(1)$ refers to a term which is irrelevant for DoF characterization}. The DoF of $m_{12}$, $m_{32}$, and $m_{13}$ is thus upper bounded by the rank of $\begin{bmatrix}\D_2\\\D_3\end{bmatrix}$ which is $\min\{N,M_2+M_3\}$ almost surely, thus
\begin{align}
\label{GenieDof2}
d_{12}+d_{32}+d_{13}\leq\min\{N,M_2+M_3\}.
\end{align}

Now, we try to enhance receiver 1 in such a way that allows it to decode $m_{23}$. To do this, we will give $\y_r^n$ and $m_{32}$ to receiver 1. Thus, after decoding $m_{21}$ and $m_{31}$, receiver 1 will have the observation $(\y_r^n, m_{31}, m_{32})$. Note that receiver 3 can decode its desired messages from $(\y_3^n,m_{31},m_{32})$ which is a degraded version of $(\y_r^n, m_{31}, m_{32})$. Thus, the enhanced receiver 1 is stronger than receiver 3 and thus, is able to decode $m_{23}$. This leads to the following bound
\begin{align}
\label{GenieBound1}
&n(R_{21}+R_{23}+R_{31}-\varepsilon_n)\nonumber\\
&\quad \leq I(m_{21},m_{31},m_{23};\y_r^n,\y_1^n,\y_3^n|m_{12},m_{13},m_{32}),
\end{align}
By using standard information-theoretic operations, we can upper bound \eqref{GenieBound1} by
\begin{align}
&n(R_{21}+R_{23}+R_{31}-\varepsilon_n)\nonumber\\
&\quad \leq \sum_{i=1}^nh\left(\left[\H_2\ \H_3\right]\begin{bmatrix}\x_{2}(i)\\\x_{3}(i)\end{bmatrix}+\z_{r}(i)\right)+n\mathcal{O}(1).
\end{align}
The DoF of these messages, i.e., $m_{21}$, $m_{31}$, and $m_{32}$, is upper bounded by the rank of $\left[\H_2\ \H_3\right]$ which leads to
\begin{align}
\label{GenieDof1}
d_{21}+d_{23}+d_{31}\leq \min\{N,M_2+M_3\}.
\end{align}
By combining \eqref{GenieDof2} and \eqref{GenieDof1} we get
\begin{align}
\label{GenieAidedDoFBound}
d_\Sigma\leq 2\min\{N,M_2+M_3\}.
\end{align}

Now, by using simple steps we can show that the bounds \eqref{CutSetDoFBound} and \eqref{GenieAidedDoFBound} can be combined into the DoF expression in Theorem \ref{Thm:DoF}.

\section{Achievability}
\label{Sec:LB}

The achievability proof of Theorem \ref{Thm:DoF} is split into 3 cases based on the dominant term in the $\min$ expression in this theorem. In some case in the proof, we will reduce the number of antennas at the relay (by ignoring some antennas) in order to guarantee the existence of an intersection subspace between the users that can be used for network coding on-the-fly in the uplink (signal space alignment for network coding \cite{LeeLimChun}).

\vspace{-.3cm}
\subsection{$d_\Sigma=2M_2+2M_3$}
\label{Sec:LB_1}
{In this case, we have $M_2+M_3\leq \min\{N,M_1\}$}, and the DoF, given by $2M_2+2M_3$, is achievable as follows. We use only $\bar{N}=M_2+M_3$ antennas at the relay. We denote the channel matrix from user $j$ to the relay by $\bar{\H}_j\in\mathbb{R}^{\bar{N}\times M_1}$ and from the relay to user $j$ by $\bar{\D}_j$. {First, the task is to find a subspace over which users 1 and 2 can align their signals in order to establish network coding on-the-fly in the uplink.} Note that $M_1\geq \bar{N}$, therefore, the columns of $\bar{\H}_1$ span the whole $\bar{N}$-dimensional space. Thus, $$\spn(\bar{\H}_2)\subset\spn(\bar{\H}_1),$$ i.e., the subspaces spanned by the columns of $\bar{\H}_1$ and $\bar{\H}_2$ intersect in an $M_2$-dimensional space spanned by the columns of $\bar{\H}_2$. This subspace is used for bidirectional communication between users 1 and 2.
Similarly, $$\spn(\bar{\H}_3)\subset\spn(\bar{\H}_1),$$ i.e., $\spn(\bar{\H}_1)$ and $\spn(\bar{\H}_3)$ intersect in an $M_3$-dimensional subspace spanned by the columns of $\bar{\H}_3$. This subspace is used for bidirectional communication between users 1 and 3.
{Note that the matrix $[\bar{\H}_2\ \bar{\H}_2]$ has linearly independent columns almost surely since $M_2+M_3=\bar{N}$ and the channels are generated randomly and independently.} The transmit signals of the users in the uplink are given by
\begin{align*}
\x_{1}=\V_{12}\u_{12}+\V_{13}\u_{13},\quad \x_{2}=\V_{21}\u_{21},\quad \x_{3}=\V_{31}\u_{31},
\end{align*}
where $\u_{12}$ is an $M_2\times1$ vector of symbols to be delivered from user 1 to user 2, and $\u_{21}$ is an $M_2\times1$ vector of symbols to be delivered from user 2 to user 1. Similarly, $\u_{13}$ and $\u_{31}$ be $M_3\times 1$ vectors to be delivered from user 1 to 3 and vice versa, respectively. The beamforming matrices $\V_{12}$, $\V_{13}$, $\V_{21}$, and $\V_{31}$ are $M_1\times M_2$, $M_1\times M_3$, $M_2\times M_2$, and $M_3\times M_3$, respectively. The relay receives
\begin{align}
\y_r&=\bar{\H}_1\V_{12}\u_{12}+\bar{\H}_2\V_{21}\u_{21}\nonumber\\
&\quad +\bar{\H}_1\V_{13}\u_{13}+\bar{\H}_3\V_{31}\u_{31}+\z_r.
\end{align}
{For the purpose of network coding, we choose the beamforming matrices such that }
\begin{align}
\spn(\bar{\H}_1\V_{12})&=\spn(\bar{\H}_2\V_{21})=\spn(\bar{\H}_2)\\
\spn(\bar{\H}_1\V_{13})&=\spn(\bar{\H}_3\V_{31})=\spn(\bar{\H}_3),
\end{align}
and both $\bar{\H}_2\V_{21}$ and $\bar{\H}_3\V_{31}$ have full column rank. {Using this design of $\V_{jk}$, we guarantee that the bidirectional communication takes place in two linearly independent subspaces.} Let $\N_{12}$ be an $M_2\times \bar{N}$ matrix whose rows span the null space of $\bar{\H}_3$, i.e., $\N_{12}\bar{\H}_3=\mathbf{0}$. The relay projects $\y_r$ onto the null space of $\bar{\H}_3$ to eliminate the contribution of $\u_{13}$ and $\u_{31}$ in $\y_r$. Denote $\N_{12}\y_r$ by $\w_{12}$. Thus the relay obtains
\begin{align*}
\w_{12}=\N_{12}\bar{\H}_1\V_{12}\u_{12}+\N_{12}\bar{\H}_2\V_{21}\u_{21}+\N_{12}\z_r.
\end{align*}
Note that since $\N_{12}$ is constructed independently of $\bar{\H}_2$, then $\N_{12}\bar{\H}_1\V_{12}$ and $\N_{12}\bar{\H}_2\V_{21}$ have full rank. Hence the relay receives a (noisy) sum of $M_2$ linearly independent observations of $\u_{12}$ and $M_2$ linearly independent observations of $\u_{21}$. $\w_{12}$ is to be sent to users 1 and 2 in the downlink. Similarly, the relay obtains $\w_{13}=\N_{13}\y_r$ by projecting $\y_r$ onto the null space of $\bar{\H}_2$ using $\N_{13}\in\mathbb{R}^{M_3\times \bar{N}}$. $\w_{13}$ is to be sent to users 1 and 3 in the downlink.

In the downlink, the relay sends $\w_{12}$ in the null space of $\bar{\D}_3$ (which spans an $M_2$-dimensional space since $\bar{N}=M_2+M_3$) and sends $\w_{13}$ in the null space of $\bar{\D}_2$ (which spans an $M_3$-dimensional space). That is,
\begin{align}
\x_r=\T_{12}\w_{12}+\T_{13}\w_{13},
\end{align}
{where $\T_{12}$ and $\T_{13}$ are projection matrices to the null spaces of $\bar{\D}_{3}$ and $\bar{\D}_{2}$, of dimensions $\bar{N}\times M_2$ and $\bar{N}\times M_3$, respectively.} User 1 receives
\begin{align}
\label{LB_Y1}
\y_1=\bar{\D}_1[\T_{12}\ \T_{13}]\begin{bmatrix}\w_{12}\\\w_{13}\end{bmatrix}+\z_1.
\end{align}
After removing the contribution of $\u_{12}$ and $\u_{13}$, and since $[\T_{12}\ \T_{13}]$ is of full rank almost surely, user 1 can decode $\u_{21}$ and $\u_{31}$ achieving $M_2+M_3$ DoF. User 2 can similarly recover $\u_{12}$ from its received signal $\y_2=\bar{\D}_2\T_{12}\w_{12}+\z_2$ achieving $M_2$ DoF, and user 3 can recover $\u_{13}$ from $\y_3=\bar{\D}_3\T_{13}\w_{13}+\z_3$ achieving $M_3$ DoF. Thus, this achieves a total of $2M_2+2M_3$ DoF equal to the upper bound.

\vspace{-.3cm}
\subsection{$d_\Sigma=M_1+M_2+M_3$}
\label{Sec:LB_2}
This case occurs if $M_1\leq M_2+M_3\leq 2N-M_1$. The transmit strategy in this case is similar to the one in \cite{LeeLimChun}, except for a different allocation of DoF between the users. 

Let us use only $\bar{N}=\nicefrac{(M_1+M_2+M_3)}{2}$ antennas\footnote{If $\bar{N}$ is not an integer, then we use 2 symbol extensions to make it integer, and proceed with designing the transmit strategy.} at the relay and denote the resulting channel matrices by $\bar{\H}_j$. Note that in this case in the uplink, the intersection subspace $\spn(\bar{\H}_1)\cap\spn(\bar{\H}_2)$ is a $d_{12}$-dimensional subspace where $$d_{12}=\nicefrac{(M_1+M_2-M_3)}{2}.$$ Similarly, $\spn(\bar{\H}_1)\cap\spn(\bar{\H}_3)$ and $\spn(\bar{\H}_2)\cap\spn(\bar{\H}_3)$ are $d_{13}$ and $d_{23}$-dimensional subspaces with
\begin{align*}
d_{13}=\nicefrac{(M_1+M_3-M_2)}{2},\quad\quad
d_{23}=\nicefrac{(M_2+M_3-M_1)}{2}.
\end{align*}
Since $d_{12}+d_{13}+d_{23}=\bar{N}$, the intersection subspaces are linearly independent almost surely. Now, similar to Section \ref{Sec:LB_1}, users 1 and 2 exchange $\u_{12}$ and $\u_{21}$ over $\spn(\bar{\H}_1)\cap\spn(\bar{\H}_2)$, and users 1 and 3 exchange $\u_{13}$ and $\u_{31}$ over $\spn(\bar{\H}_1)\cap\spn(\bar{\H}_3)$. Additionally, users 2 and 3 exchange $\u_{23}$ and $\u_{32}$ over $\spn(\bar{\H}_2)\cap\spn(\bar{\H}_3)$. Let the transmit signals be
\begin{align}
\x_{1}&=\V_{12}\u_{12}+\V_{13}\u_{13}\\
\x_{2}&=\V_{21}\u_{21}+\V_{23}\u_{23}\\
\x_{3}&=\V_{31}\u_{31}+\V_{32}\u_{32},
\end{align}
where the information vectors $\u_{12}$ and $\u_{21}$ are $d_{12}\times1$, $\u_{13}$ and $\u_{31}$ are $d_{13}\times1$, and $\u_{23}$ and $\u_{32}$ are $d_{23}\times1$.
The beamforming matrices $\V_{12}$, $\V_{13}$, $\V_{21}$, $\V_{23}$, $\V_{31}$, and $\V_{32}$ are $M_1\times d_{12}$, $M_1\times d_{13}$, $M_2\times d_{12}$, $M_2\times d_{23}$, $M_3\times d_{13}$, and $M_3\times d_{23}$, respectively. Since the intersection subspaces are linearly independent almost surely, the relay can obtain
\begin{align}
\w_{12}&=\N_{12}\bar{\H}_1\V_{12}\u_{12}+\N_{12}\bar{\H}_2\V_{21}\u_{21}+\N_{12}\z_{r}\\
\w_{13}&=\N_{13}\bar{\H}_1\V_{13}\u_{13}+\N_{13}\bar{\H}_3\V_{31}\u_{31}+\N_{13}\z_{r}\\
\w_{23}&=\N_{23}\bar{\H}_2\V_{23}\u_{23}+\N_{23}\bar{\H}_3\V_{32}\u_{32}+\N_{23}\z_{r},
\end{align}
where $\N_{12}$, $\N_{13}$, and $\N_{23}$ are the suitable zero-forcing matrices for eliminating the undesired variables, of dimensions $d_{12}\times \bar{N}$, $d_{13}\times \bar{N}$,and $d_{23}\times \bar{N}$, respectively. The relay sends 
\begin{align}
\x_r=\T_{12}\w_{12}+\T_{13}\w_{13}+\T_{23}\w_{23}.
\end{align}
where $\T_{12}$, $\T_{13}$, and $\T_{23}$ are zero-forcing beamforming matrices of dimensions $\bar{N}\times d_{12}$, $\bar{N}\times d_{13}$, and $\bar{N}\times d_{23}$, respectively. That is, 
\begin{align}
\bar{\D}_3\T_{12}=\mathbf{0},\quad \bar{\D}_2\T_{13}=\mathbf{0},\quad\bar{\D}_1\T_{23}=\mathbf{0}. 
\end{align}
The received signal at user 1 is similar to \eqref{LB_Y1}. Since $[\T_{12}\ \T_{13}]$ is of full column rank almost surely ($d_{12}+d_{13}=M_1<\bar{N}$), then user 1 can decode $\u_{21}$ and $\u_{31}$ achieving $d_{12}+d_{13}$ DoF. User 2 can similarly achieve $d_{12}+d_{23}$ DoF, and user 3 can achieve $d_{13}+d_{23}$ DoF. In total, this scheme achieves $M_1+M_2+M_3$ DoF equal to the upper bound.

\vspace{-.3cm}
\subsection{$d_\Sigma=2N$}

This case occurs if $N<\min\{M_2+M_3,\nicefrac{(M_1+M_2+M_3)}{2}\}$. In this case, we use only $\bar{M}_j\leq M_j$ antennas at each user such that $2N=\min\{2\bar{M}_2+2\bar{M}_3,\bar{M}_1+\bar{M}_2+\bar{M}_3\}$. Then, in order to achieve $2N$ DoF, we use the same scheme as in Sections \ref{Sec:LB_1} and \ref{Sec:LB_2}.
This completes the proof of the achievability of Theorem \ref{Thm:DoF}.

\vspace{-.3cm}
\section{Discussion}
\label{Sec:Disc}
In this section, we give an easy constructive method for designing the pre-coding and post-coding matrices of the strategy in \ref{Sec:LB}. This is done using a graphical illustration and an example. The presented graphical illustration has the advantage that it can be applied for higher than 3 dimensions, in contrast to the more common 3-dimensional graphical signal-space representations which is limited to 3 dimensions.

\begin{figure}
\centering
\includegraphics[width=.750\columnwidth]{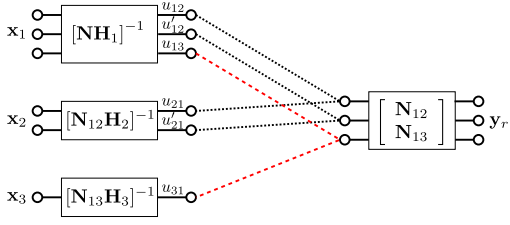}
\caption{The uplink in a Y-channel with $(M_1,M_2,M_3,N)=(3,2,1,3)$ showing the pre-coding and post-coding matrices. $\N_{12}$ is orthogonal to $\H_3$, $\N_{13}$ is orthogonal to $\H_2$, and $\N=[\N_{12}^\dagger,\ \N_{13}^\dagger]^\dagger$. Note the desirable structure of the inner channel resulting after pre-coding and post-coding. The relay obtains noisy observations of $x_{12}+x_{21}$, $x_{12}'+x_{21}'$, and $x_{13}+x_{31}$.}
\label{Fig:Illustration_U}
\end{figure}

\begin{figure}
\centering
\includegraphics[width=.85\columnwidth]{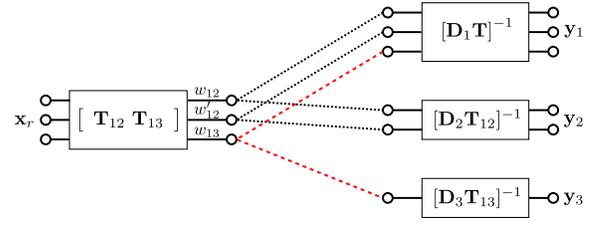}
\caption{The downlink in a Y-channel with $(M_1,M_2,M_3,N)=(3,2,1,3)$ showing the pre-coding and post-coding matrices. $\T_{12}$ and $\T_{13}$ are orthogonal to $\D_3$ and $\D_2$, respectively, $\T=[\T_{12},\  \T_{13}]$, and $w_{12}$, $w_{12}'$, and $w_{13}$ represent noisy linear combinations of $(x_{12},x_{21})$, $(x_{12}',x_{21}')$, and $(x_{13},x_{31})$, respectively.}
\label{Fig:Illustration_D}
\end{figure}

Consider a Y-channel with $(M_1,M_2,M_3,N)=(3,2,1,3)$. This Y-channel has $2M_2+2M_3=6$ DoF according to Theorem \ref{Thm:DoF}. We transform the channel into two sub-channels, one dedicated for bidirectional communication between users 1 and 2, and one for users 1 and 3 (see Fig. \ref{Fig:Illustration_U}). The received signal at the relay in the uplink is
\begin{align}
\y_r=\H_1\x_1+\H_2\x_2+\H_3\x_3+\z_r.
\end{align}
We need to construct a post-coding matrix $\N\in\mathbb{R}^{3\times3}$ at the relay, which zero-forces interference from user 3 in the sub-channel for bidirectional communication between users 1 and 2. Thus, let us choose the first two rows of $\N$ as a matrix spanning the null space of $\H_3$, denoted $\N_{12}\in\mathbb{R}^{2\times3}$, i.e., $\N_{12}\H_3=\mathbf{0}$. Similarly, the last row of $\N$ is chosen as $\N_{13}\in\mathbb{R}^{1\times3}$ such that $\N_{13}\H_2=\mathbf{0}$. Thus, we obtain the post-coding matrix $\N=[\N_{12}^\dagger\ \N_{13}^\dagger]^\dagger$. Then, after post-coding, the relay has the signal $\N\y_r$ given by
\begin{align*}
\begin{bmatrix}\N_{12}\H_1\\\N_{13}\H_1\end{bmatrix}\x_1+\begin{bmatrix}\N_{12}\H_2\\0\end{bmatrix}\x_2+\begin{bmatrix}\mathbf{0}\\\N_{13}\H_3\end{bmatrix}\x_3+\N\z_r.
\end{align*}
Note that this establishes a $2\times2$ MIMO sub-channel between user 2 and the first two components of $\N\y_r$ without causing interference at the third component. Similarly for user 3, a SISO channel is established which does interfere with the first two components of $\N\y_r$. Now the task is to construct the pre-coding matrices such that the signal from user 1 to user 2, $[u_{12},\ u_{12}']^\dagger$, is only seen at the first 2 components of $\N\y_r$, and the signal from user 1 to user 3, $u_{13}$ is only seen at the last component of $\N\y_r$. This can be accomplished by choosing a pre-coding matrix at user 1 which diagonalizes the channel to $\N\y_r$. Users 1, 2, and 3 send 
\begin{align*}
\x_1=\V_1\begin{bmatrix}u_{12}\\u_{12}'\\ u_{13}\end{bmatrix},\quad \x_2=\V_2\begin{bmatrix}u_{21}\\ u_{21}'\end{bmatrix},\quad \x_3=\V_3u_{31},
\end{align*}
where $\V_1=[\N\H_1]^{-1}$, $\V_2=[\N_{12}\H_2]^{-1}$, and $\V_3=[\N_{13}\H_3]^{-1}$. The resulting received signal at the relay is then
\begin{align*}
\begin{bmatrix}w_{12}\\w_{12}'\\w_{13}\end{bmatrix}=\N\y_r=\begin{bmatrix}u_{12}+u_{21}\\u_{12}'+u_{21}'\\u_{13}+u_{31}\end{bmatrix}+\N\z_r.
\end{align*}
Now, we use a similar strategy in the downlink to obtain the channel structure shown in Fig. \ref{Fig:Illustration_D}. The relay sends $\w=[\w_{12}^\dagger,\ \w_{13}^\dagger]^\dagger$ where $\w_{12}=[w_{12}, w_{12}']^\dagger$, and $\w_{13}=w_{13}$ along the columns of $\T=[\T_{12}\ \T_{13}]$ where $\T_{12}\D_3=\vec{0}$ and $\T_{13}\D_2=\vec{0}$. User 1 uses a post-coding matrix $[\D_1\T]^{-1}$, user 2 uses a post-coding matrix $[\D_2\T_{12}]^{-1}$, and user 3 uses a post-coding matrix $[\D_3\T_{13}]^{-1}$. As a result, the received signal at the three users become
\begin{align*}
[\D_1\T]^{-1}\y_1&=\w+[\D_1\T]^{-1}\z_1\\
[\D_2\T_{12}]^{-1}\y_2&=\w_{12}+[\D_2\T_{12}]^{-1}\z_2\\
[\D_3\T_{13}]^{-1}\y_3&=\w_{13}+[\D_3\T_{13}]^{-1}\z_3,
\end{align*}
from which each user can recover its desired signal.


\vspace{-.2cm}
\bibliography{myBib}

\end{document}